\documentclass[%
 reprint,
 amsmath, 
 aps, prl
]{revtex4-2}

\usepackage{graphicx}
\usepackage{dcolumn}
\usepackage{bm}
\usepackage{amsmath}
\usepackage{unicode-math}
\usepackage{physics}
\usepackage[dvipsnames]{xcolor}

\begin{document}

\preprint{APS/123-QED}

\title{Evidence for cavity-induced metallic phase from entropic electron correlation effects.}
\author{Jacob Horak}%
\affiliation{Max Planck Institute for the Structure and Dynamics of Matter and Center for Free-Electron Laser Science, Luruper Chaussee 149, Hamburg 22761, Germany}
\affiliation{The Hamburg Center for Ultrafast Imaging, Luruper Chaussee 149, 22761 Hamburg, Germany}

\author{Dominik Sidler}
 \affiliation{Zurich University of Applied Sciences, School of Engineering, 
 Technikumstrasse 71, Winterthur 8400, Switzerland}
\affiliation{%
 Max Planck Institute for the Structure and Dynamics of Matter and Center for Free-Electron Laser Science, Luruper Chaussee 149, Hamburg 22761, Germany
}
 \email{dominik.sidler@zhaw.ch}
 

%



\date{\today}

\begin{abstract}
 Experiments have revealed that collective strong coupling in optical cavities can  drastically alter the conductivity and dielectric properties of molecular ensembles, or even trigger abrupt phase transitions (in Rayleigh scattering or dispersion force driven conformational equilibrium).
 Until now, the underlying physical mechanism has remained unknown. A recently proposed theory attributes these collective effects to cavity-induced electron correlations that share properties of the known spherical Sherrington-Kirkpatrick model of a spin glass. As a consequence, entropic effects can trigger a phase transition in the inter-molecular electron correlations that could potentially explain the above experiments. In the following work, we derive analytic expressions for the correlation (free-)energy-induced polarizability changes, which reveal that under certain conditions collective strong coupling can even render our molecular ensemble metallic by entropically ionizing collectively degenerate molecular orbitals. Our results are a first major step towards a holistic theory of cavity-mediated electron correlation effects.
\end{abstract}

\maketitle

The exciting field of polaritonic chemistry emerged when experiments showed chemical changes due to the changed electromagnetic environment inside an optical cavity. \cite{ebbesen_hybrid_2016, thomas_groundstate_2016, thomas_tilting_2019,hertzog_strong_2019,ruggenthaler_quantum-electrodynamical_2018,sidler_perspective_2022,fregoni_theoretical_2022,ebbesen_introduction_2023,ruggenthaler_understanding_2023,bhuyan_rise_2023,hirai_molecular_2023,simpkins_control_2023,mandal_theoretical_2023,xiang_molecular_2024} Remarkably, these chemical changes typically occur without external illumination. 
Traditionally, the chemical changes are attributed and explained by the formation of polaritons, hybrid particles made up of molecular excitations strongly coupled to cavity photon modes and separated by an energy called vacuum Rabi splitting. However, most collective vibrational strong coupling (VSC) cavities are tuned to IR modes that only scarcely get populated at room temperature, thus rather suggesting electronic / vibrational ground state modifications should cause these effects that are not (directly) captured by polaritons. Aside from ground state modifications, another theoretical issue is the so called large-$N$ problem emergent from simplified collective models. It states that significant local (molecular) chemical changes are implausible, due to vanishing local light-matter couplings,\cite{martinez-martinez_can_2018,schwennicke_when_2025} or remain finite\cite{sidler_polaritonic_2021,sidler_unraveling_2024, castagnola_changes_2024, horak_analytic_2025,castagnola_realistic_2025}, but likely too small to explain the majority of experiments.  

Recent theoretical advances show that an energetic collective picture is incomplete and that cavity-mediated electron correlation effects (i.e., the Pauli-principle) offer a collective entropic mechanism to change chemistry locally, provided one relaxes the commonly applied dilute gas assumption.\cite{sidler_collectively-modified_2026,sidler_cavity-mediated_2026} This theoretical advancement was possibly by mapping the transverse electron problem to the spherical Sherrington-Kirkpatrick model, known from spin glass physics. Its analytic solution suggests the emergence of three different phases for collective dispersion effects / electron correlations. First, the absence of transversal electron correlations (only bare matter Coulombic correlations). Second, a para correlated phase and third a spin glass correlated phase (suggesting aging effects and break down of standard fluctuation dissipation relations).\cite{sidler_cavity-mediated_2026} Theory suggests that collective correlations are entropically stabilized, i.e., a correlation free-energy lowering can favor local orbital excitations (i.e., local chemical changes).\cite{sidler_cavity-mediated_2026} 
From an experimental side, temperature-dependent  phase (-like) transitions and cavity-modified London dispersion forces (electron correlations) have indeed been reported for different experimental observables, which support the theory of collective electron correlation  phases.\cite{sidler_collectively-modified_2026,sidler_cavity-mediated_2026} Notice that recently there has also been a macroscopic condensation theory proposed to potentially explain collectively induced chemical changes \cite{mondal_macroscopic_2026}. Still there are many open questions left and we are at a very early stage to connect the spin glass theory to experiments. One of the major difficulty is the calculation of chemical observables, since the spin glass solution is infinitely degenerate (thermodynamic limit) and thus requires a fundamental non-perturbative discussion of chemical observables. In the following, we do a first major step towards its understanding by deriving analytic expressions for the local polarizability changes due to glassy electron correlations. Our subsequent calculations also suggest that the collective electron correlations can under certain conditions effectively turn the molecular ensemble into a metal, i.e., the ionization of (some) electrons can become entropically favorable inside the cavity. This could be the missing piece for the mechanistic understanding of various experiments that report strong changes in conductivity,\cite{kumar_extraordinary_2024,jarc_cavity-mediated_2023} dielectric constant,\cite{fukushima_inherent_2022} or Rayleigh scattering\cite{sandeep_cluster_2026}.  
The manuscript is structured as follows, first we briefly recapitulate the essentials of the spin glass mapping starting from the non-relativistic Pauli-Fierz theory.  Afterwards, we shortly recapitulate the emergence of different electron correlation phases, before we continue with the non-perturbative derivation of entropically induced polarizability changes and their connection to known experiments.

Within the cavity-Born-Oppenheimer partitioning, the electronic Hamiltonian is given as,\cite{ruggenthaler_understanding_2023}
\begin{widetext}
\begin{align}
%
\hat{H}^{e} &=    \sum_{i=1}^{2N}\bigg\{\frac{\hat{\vec{p}}_i^2}{2m} +V_{\rm Coul}(\hat{\vec{r}}_i,\textit{\textbf{R}})-\vec{\lambda} \cdot e\hat{\vec{r}}_i \bigg(\vec{\lambda}\cdot \sum_{j=1}^{N_{\rm nuk}}Z_je\vec{R}_j-\omega_\beta  q_\beta\bigg)\bigg\}\nonumber\\
&
+\underbrace{\frac{1}{2}\sum_{i=1}^{2N}\sum_{j=1,j\neq i}^{2 N}\frac{e^2}{4\pi \epsilon_0|\hat{\vec{r}}_i - \hat{\vec{r}}_j|}}_{\hat{W}_\parallel}+\underbrace{\frac{\big(\vec{\lambda} \cdot \sum_{i=1}^{2N}e\hat{\vec{r}}_i\big)^2}{2} }_{\hat{W}_\perp}.
\label{eq:H}
\end{align}
\end{widetext}
The first line contains solely one-electron terms, with kinetic energy, Coulomb attraction of the nuclei $V_{\rm Coul}$ and the linear coupling to a single effective cavity mode $q_\beta$ of frequency $\omega_\beta$, linear polarization $\vec{\lambda}/|\lambda|$ and light-matter coupling parameter $\lambda=|\vec{\lambda}|$. Electronic position operators are indicated by $\hat{\vec{r}}_i$ with elementary charge $e$. The number of electrons is $2N$ and the positive nuclei enter with $\vec{R}_j$ and charge $Z_j e$. The second line, contains longitudinal (Coulomb) two-electron operators $\hat{W}_\parallel$ and the transversal dipole self-energy $\hat{W}_\perp$.

To obtain insight into the transversal electron correlations emerging from our degenerate subspace, we use a single Hartree-Fock  reference state  $|\Phi_0 \rangle$  and then perform a full configuration interaction (FCI) expansion on the degenerate subspace.
As shown in Ref. \cite{sidler_cavity-mediated_2026}, by assuming a uniform ensemble, i.e., equal probability weight per FCI excited state on the degenerate subspace, collective transversal correlations can be approximated by configuration interaction singles (CIS) as follows
\begin{eqnarray}
    E_{\rm corr}\approx\langle \hat{N}_v\rangle_{\rm uniform} E_{\rm CIS,corr}\label{eq:corr_approx}
\end{eqnarray}
with
\begin{eqnarray}
    \langle \hat{N}_v\rangle_{\rm uniform}=\frac{N_o N_u}{N_o^{\rm tot}}.\label{eq:nv_uniform}
\end{eqnarray}
The numbers of occupied and unoccupied states are fixed by $N_o, N_u$, with total degeneracy $N_o^{\rm tot}=N_o+N_u$. The CIS ansatz wavefunction is given by $|\Phi^\mathrm{CIS}\rangle
= s_0|\Phi_0 \rangle
  + \sum_{m\in N_o}\!\sum_{p\in N_u} s_m^p\,|\Phi_m^p\rangle$.

\begin{figure}
    \centering
    \includegraphics[width=1\linewidth]{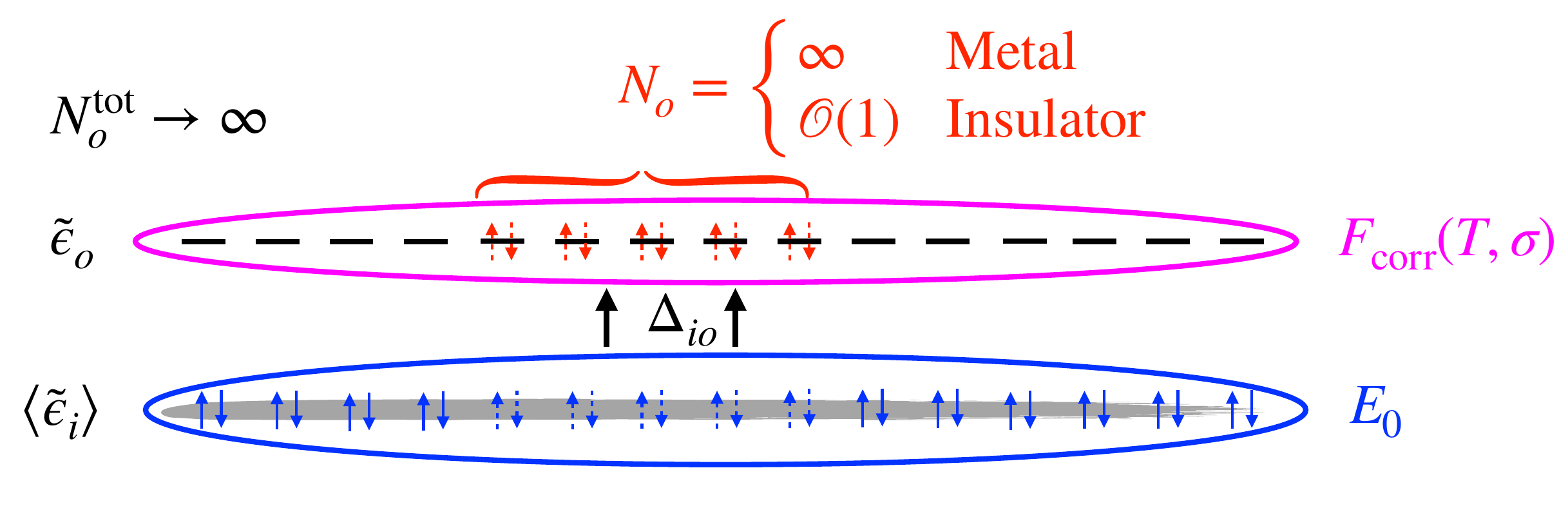}
    \caption{Simplified collective 2-level model to qualitatively investigate the formation of metallic transversal electron correlation phases under collective SC in the thermodynamic limit, as introduced in Ref. \citenum{sidler_cavity-mediated_2026}. Mean Hartree-Fock orbital energies are labeled $\langle\tilde{\epsilon}_i\rangle $ which jointly accumulate to the ensemble Hartree-Fock energy $E_0$ (blue). The cavity-induced $N_o^{\rm tot}$-fold degenerate space  is occupied $N_o$-times with excitation energy  $\Delta_{io}=\tilde{\epsilon}_o-\langle\tilde{\epsilon}_i\rangle>0$. Depending on the chemical ensemble, exciting $N_o$ orbitals can lower the transversal correlation free-energy $F_{\rm corr}$ entropically, and thus becoming energetically favorable. Two qualitatively different cases can be distinguished in the thermodynamic limit: The metallic case with a divergent occupation number vs. the insulating case with finite occupation number.\cite{sidler_cavity-mediated_2026} }
    \label{fig:setup}
\end{figure}

As recently shown in Ref.~\cite{sidler_collectively-modified_2026}, the resulting CIS transversal correlation energy problem can be mapped onto the spherical Sherrington-Kirkpatrick model of a spin glass.  Therefore, by using Eq. \eqref{eq:corr_approx}, the  transversal FCI correlation energy minimization problem on the degenerate  subspace has the following form \cite{sidler_collectively-modified_2026}
\begin{equation}
E_{\rm corr} =- \sum_{i< j}^{N_{\rm deg}} J_{ij}\,\sigma_i\sigma_j,
  \qquad \sum_{i=1}^{N_{\rm deg}}\sigma_i^2 = 1
\label{eq:ECIS}
\end{equation}
where   $\sigma_i$ are normalized CIS amplitudes emergent from $s_m^p\mapsto \sigma_i$. They play the role of continuous spin variables that are normalized to one. The random interactions $J_{ij}$ emerge from two-electron integrals of the dipole-self energy, i.e., from $\hat{W}_\perp$, in combination with random molecular orientation (rotation) with respect to the linear cavity-polarization~\cite{sidler_collectively-modified_2026}. More specifically
$
    {\rm Var}(J_{ij}) = \langle \hat{N}_v\rangle_{\rm uniform}^2\lambda^4 e^4{\rm Var}(\bra{\phi_m}\hat{z}\ket{\phi_n}\bra{\phi_q}\hat{z}\ket{\phi_p}),\label{eq:sigma}
$
assuming a cavity polarization along $z$-axis.
The spin glass problem in Eq. \eqref{eq:ECIS} possesses an analytic solution if the random interactions are assumed to belong to the equivalence class of zero mean with extensive variance
 \begin{eqnarray}
    {\rm Var}(J_{ij})=\sigma^2_0 N_{\rm deg}. \label{eq:varjij}
 \end{eqnarray}
Hence, the transversal correlation minimization problem $E_{\rm corr}$ is equivalent to the spherical Sherrington-Kirkpatrick (SSK) model of a spin glass.~\cite{sherrington_solvable_1975,baik_spherical_2021} Its analytic solution for  the  free energy   is given as $F_{N_{\rm deg}}=-k_B T \log(Z_{N_{\rm deg}})=N_{\rm deg} f$, with free energy per spin,
\begin{widetext}
\begin{eqnarray}
    f(T,\sigma_0)
    =&\begin{cases}
        -  \frac{\sigma_0^2}{4 k_B T}, & \sigma_0\leq k_B T \\
        -\sigma_0 +\frac{k_B T}{2}\log{\big(\frac{\sigma_0 }{k_B T}\big)}+\frac{3 k_B T}{4}, &\sigma_0> k_B T
    \end{cases}
    \label{eq:freeenergy}
\end{eqnarray}
\end{widetext}
and critical temperature $T_c=\sigma_0/k_B$. Notice, when deriving Eq.~\eqref{eq:freeenergy}, the thermodynamic limit is implicitly assumed, i.e., $N_{\rm deg}\rightarrow\infty$. 
As derived in Ref. \cite{sidler_cavity-mediated_2026}, the required extensive scaling of random interactions for the SSK thermodynamic limit ${\rm Var}(J_{ij})\sim N_{\rm deg}\rightarrow\infty$ can be reached by two distinct microscopic mechanisms. Either, the random interactions  scale extensively already on CIS level with   $\langle\hat{N}_v\rangle_{\rm uniform}\sim \mathcal{O}(1)$, or due to half-filling with $\langle\hat{N}_v\rangle_{\rm uniform}\sim \mathcal{O}(N_{\rm deg})$ . The first case suggests \textit{insulating behavior} due to an (almost) empty or (almost) fully occupied degenerate collective state and the second, fractionally filled,
suggests a (rather) \textit{metallic behavior}, i.e., a high mobility of the $N_o$ occupied orbitals in the degenerate subspace (see Fig. \ref{fig:setup}). 
In the following, we aim to characterize these two different physical mechanisms by their respective static polarizability. In particular, do we really find a locally (!) diverging transversal polarizability, as one expects in a metallic phase, or does the effect remain locally finite and diverges only collectively? In both cases, we assume the bare molecular ensemble (in absence of transversal correlation effects) to be non-conducting. 

Before we proceed, we briefly recapitulate the simplified (2-level)  model, introduced in Ref.\cite{sidler_cavity-mediated_2026} (see Fig. \ref{fig:setup}), for an ensemble of molecules under collective strong coupling. It has non-degenerate Hartree-Fock orbitals with average energy $\langle\tilde{\epsilon}_i\rangle$, which can be excited to the collectively degenerate state $\tilde{\epsilon}_o$, effectively forming the SSK spin glass in the thermodynamic limit. The collective degeneracy may entropically favor local excitations, giving rise to three different phases with the following generic phase diagram\cite{sidler_cavity-mediated_2026}
\begin{widetext}
\begin{eqnarray}
    E = \begin{cases}
        E_0& {\rm if}\ f(T,\sigma_0)\geq -\Delta_{io}\\
        E_0+ \Delta E+F_{\rm corr}(T,\sigma_0), & {\rm otherwise}
    \end{cases}.
    \label{eq:correlation_spinglass}
\end{eqnarray}
\end{widetext}
The excitation from an average non-degenerate to a degenerate orbital costs the average HF energy $\Delta_{io}=|\tilde{\epsilon}_o-\langle\tilde{\epsilon}_i\rangle|$, which amounts to $\Delta E= N_o \Delta_{io}$. The HF reference energy is given by $E_{0}=\bra{\Phi_0}\hat{H}^e\ket{\Phi_0}$. For the metallic case, a collectively correlated phase is entered if $ f(T,\sigma_0)< -\Delta_{io}$ according to Eq. \eqref{eq:correlation_spinglass}.
In contrast, for the insulating case (see Fig. \ref{fig:setup}), no generic condition is currently known for entering the collectively correlated phases, i.e., whether $F_{\rm corr}(T,\sigma_0)\neq 0$ is achievable. Still, if we assume non-vanishing collective transversal correlations, we can subsequently investigate the resulting electronic polarizability change and its phase transition from para to spin glass correlations, which may help to interpret experimental results microscopically.

Based on these known results,\cite{sidler_cavity-mediated_2026} we subsequently
investigate the polarizability change induced by the collective transversal electron correlations. We, therefore, apply a constant external electric field $E_{loc}$ along the cavity polarization to our electronic Hamiltonian operator, i.e., $
    \hat{H}(E_{loc})= \hat{H}^e + \hat{H}^{'} =\hat{H}^e-eE_{loc}\sum_{i=1}^{2N}\hat{\vec{z}}_i
$.
Next, we aim to solve the perturbed Hamiltonian $\hat{H}(E_{loc})$ non-perturbatively, in order to access the impact of the external electric field on the infinitely degenerate state (thermodynamic limit) causing the cavity-induced transversal correlation phases. Note that standard perturbative approaches break down due to the degeneracy in the thermodynamic limit. Fortunately, we can solve the transverse electron correlation problem induced by $\hat{H}(E_{loc})$ non-perturbatively  using the previous CIS ansatz $|\Phi^{\rm CIS}\rangle$, thereby yielding analytic insight, as when solving $\hat{H}^e$. Eventually, one finds (see Appendix):  
\begin{equation}
E(E_{loc}) = E_{0}
  - \sum_{i< j}^{N_{\rm deg}} (J_{ij}+E_{\rm loc}L_{ij})\sigma_i\sigma_j,
  \qquad \sum_{i=1}^{N_{\rm deg}}\sigma_i^2 = 1,
\label{eq:ECIS_E}
\end{equation}
Notice both random numbers $J_{ij}$ and $L_{ij}$ are connected to the transition dipole fluctuations, however, they have different form and different physical origin. $J_{ij}$ emerges from the two-electron integrals, i.e., describes internal interaction, whereas $L_{ij}$ emerges from non-vanishing one-electron integrals caused by the applied external electric field.  For the subsequent discussion, we proceed by reasonably assuming uncorrelated random numbers $J_{ij}$ and $L_{ij}$.
We then find,
\begin{eqnarray}
     {\rm Var}(J_{ij}+E_{loc}L_{ij})=\sigma^2(E_{loc}) N_{\rm deg} ,\label{eq:varjij}
 \end{eqnarray} 
with 
\begin{eqnarray}
    \sigma^2(E_{loc})  &=&\sigma_0^2+E_{loc}^2\tau_0^2 \\
    %
    &=&  \gamma^4\tau_0^4+E_{loc}^2 \tau_0^2
    .\label{eq:sigma_E}
\end{eqnarray}
and $\tau_0^2 N_{\rm deg}=e^2 Var(\bra{\phi_m}\hat{z}\ket{\phi_n})\langle \hat{N}_v\rangle^2$.   Consequently,  Eq. \eqref{eq:ECIS_E} becomes
\begin{eqnarray}
    E(E_{loc})=E_0+F_{\rm corr}\big(T,\sigma(E_{\rm loc})\big),
\end{eqnarray}
in the thermodynamic limit. Notice, in Eq. \eqref{eq:sigma_E} we have introduced the correlation factor $\gamma$ that relates the unperturbed random fluctuations to the electric field induced ones as follows $\sigma_0^2=\gamma^4\tau_0^4$.

The cavity-induced transversal polarizability $\alpha_\perp$  can thus be calculated per spin variable $i$ from $ \alpha_\perp = -\frac{\partial^2 f(T,\sigma(E_{loc})) }{\partial^2 E_{loc}}\Big\rvert_{E_{loc}=0}$ with Eq. \eqref{eq:freeenergy}, yielding:
\begin{eqnarray}
    \alpha_\perp = \begin{cases}
        \frac{\tau_0^2}{2 k_B T}, & \quad\sigma_0\leq k_B T \\
        \frac{1}{\gamma^2}\bigg[1 - \frac{k_B T}{2  \sigma_0}\bigg], & \quad\sigma_0> k_B T 
    \end{cases}. \label{eq:polarizability}
\end{eqnarray}
Eq. \eqref{eq:polarizability} can lead to qualitative different physics/chemistry, depending on the microscopic origin of collective transversal electron correlations.

\begin{figure*}
    \centering
    \includegraphics[width=0.7\textwidth]{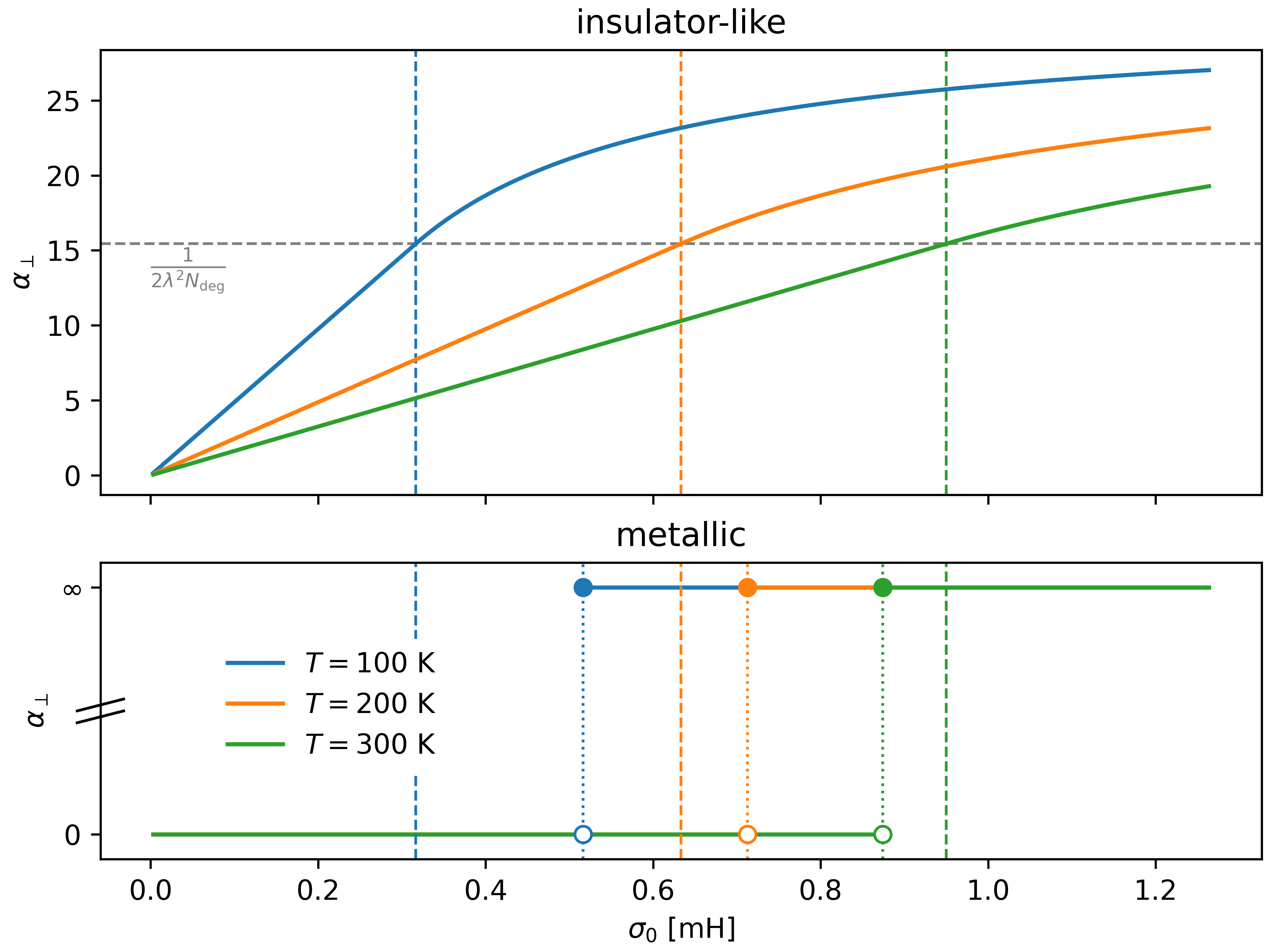}
    \caption{Polarizability from transversal correlations for insulator-like (top) and metallic regime (bottom) given by Eqs. \eqref{eq:insulator} and \eqref{eq:metallic}, respectively for three different temperature $T$ choices.
    Phase transitions are indicated by dashed vertical  lines. The critical point $\sigma_0=k_B T$ of the para- to spin glass phase transition remains the same for both cases and is indicated by long dashed lines.
    However, in the metallic case (bottom) the transversal polarizability diverges and we cannot observe the change from the para to spin glass correlated phase. Instead, we can identify the critical points for the uncorrelated to correlated phase transition. As we see from the long dashed lines, we either  enter the spin glass phase directly (blue) or we first enter the para correlated phase (orange, green). 
    The figure was created using $\Delta_{io} = 0.275$~mH  and $\lambda^2 N_\mathrm{deg} = 0.1799$. 
    }
    \label{fig:polarizabilities}
\end{figure*}

\textbf{Insulating case (assuming $\langle \hat{N}_v\rangle=1$, $E_{\rm CIS,corr}=N_{\rm deg}f$):}
For the insulating case, we know from $\langle \hat{N}_v\rangle=1$ the extensive scaling of the SSK model must arise from microscopic correlations among the fluctuations (their characterization and microscopic origin is non-trivial and will be the topic of a subsequent publication). We thus impose ${\rm Var}(\bra{\phi_m}\hat{z}\ket{\phi_n}\bra{\phi_q}\hat{z}\ket{\phi_p})={\rm Var}(\bra{\phi_m}\hat{z}\ket{\phi_n})^2 N_{\rm deg}$.
As consequence, we identify
    $\gamma^4=\lambda^4 N_{\rm deg}^2\sim \mathcal{O}(1)$, assuming the usual collective scaling of $\lambda\sim1/\sqrt{N}$. The resulting polarizability per spin, emergent from the collective transversal correlations, 
\begin{eqnarray}\label{eq:insulator}
    \alpha^{\rm insulator}_\perp =  \begin{cases}
        \frac{1}{\lambda^2 N_{\rm deg}}\frac{\sigma_0}{2 k_B T}, & \sigma_0\leq k_B T \\
        \frac{1}{\lambda^2 N_{\rm deg}}\bigg[1 - \frac{k_B T}{2  \sigma_0}\bigg], & \quad\sigma_0> k_B T,
    \end{cases}
\end{eqnarray}
is finite; thus all electrons remain localized, and the collectively correlated molecular ensemble indeed remains insulating.
Without further assumption on the microscopically correlated fluctuations, we cannot make any statement about entering the insulating collectively correlated phase. We can only qualitatively interpret the microscopic polarizability dependence on $\sigma_0$, as shown in Fig. \ref{fig:polarizabilities}. At the critical temperature $T_c=\sigma_0/k_B$, the linearly increasing polarizability of the para-correlated phase enters the spin glass correlation regime continuously, eventually saturating at finite $\alpha^{\rm insulator}_\perp \rightarrow 1/{\lambda^2 N_{\rm deg}}$.

\textbf{Metallic case (assuming $\langle \hat{N}_v\rangle=N_{\rm deg}$, $E_{\rm CIS,corr}=f$):}  From Eq. \eqref{eq:nv_uniform} we notice $\langle\hat{N}_v\rangle_{\rm uniform}\sim N_o $ 
yielding the interpretation $F=\langle\hat{N}_v\rangle_{\rm uniform} f$. This assigns the local correlation free-energy $f$ and thus polarizability $\alpha^{\rm metal}_\perp$ to each of the $N_o$ entropically excited orbitals in the collectively degenerate state $\tilde{\epsilon}_o$. With ${\rm Var}(\bra{\phi_m}\hat{z}\ket{\phi_n}\bra{\phi_q}\hat{z}\ket{\phi_p})={\rm Var}(\bra{\phi_m}\hat{z}\ket{\phi_n})^2 $, one therefore finds $\gamma^4=\lambda^4/N_{\rm deg}$.  In the thermodynamic limit, the metallic polarizability can be written as,
\begin{eqnarray}\label{eq:metallic}
    \alpha^{\rm metal}_\perp =  
    \begin{cases}
        0& {\rm if}\ f(T,\sigma_0)\geq -\Delta_{io}\\
        \infty
    \end{cases},
\end{eqnarray}
with  vanishing $\gamma\rightarrow 0$  for finite  $\lambda^2>0$ and divergent $\tau_0\rightarrow\infty$, assuming finite $\sigma_0>0$. A divergent local polarizability indicates that our initially insulating (bare) molecular ensemble can indeed become metallic for finite fractional occupation at $f(T,\sigma_0)< -\Delta_{io}$ if either the para- or spin glass correlated phase is entered (see Fig. \ref{fig:polarizabilities}). This suggests the following interpretation. For fractionally filling, a divergent number $\langle\hat{N}_v\rangle\sim N_o$ of  occupied orbitals effectively acts as free electrons, and thus entropically ionized (excited) by the free-energy reduction of the transversal electron correlations. In our model, both collectively correlated phases (para and spin glass correlated) show the same metallic behavior at $T_c$ and cannot be further distinguished from static polarizability measurements. To overcome this limitation (e.g., measure spin glass aging effects), a time-resolved picture is required, which goes beyond the scope of this work.  Furthermore, we cannot make any statement about the number density of these collectively ionized orbitals. Most likely its number density remains  low for most practical chemical setups. However, this phase transition should appear in chemical observables that are sensitive to (few) ionized electrons, whereas average chemical properties are most likely only marginally affected. Indeed, there is different experimental evidence that collective strong coupling can cause such effects. For example, experiments suggest that the static dielectric constant of liquid water under VSC may exceed that of ice, depending on the collective coupling strength.\cite{fukushima_inherent_2022} Typically, those dielectric measurements in water are very sensitive to dissolved ions and thus may also sensitively react to cavity-mediated entropic ionization. A cavity-induced metallic behavior also nicely connects to the experiments of Kumar et al. \cite{kumar_extraordinary_2024}, which show
 an enhancement of the electrical conductivity by six orders of magnitude for intrinsically nonconducting  polymers such as polystyrene, deuterated polystyrene, and poly(benzyl methacrylate) under VSC.\cite{kumar_extraordinary_2024}
The emergence of abrupt cavity-induced phase transitions under VSC  have  been directly observed  using NMR\cite{patrahau_direct_2024} as well as Rayleigh-scattering experiments.\cite{sandeep_cluster_2026} In the scattering experiments,  an abrupt, two-order of magnitude increase of the scattering intensities has been observed, at a temperature-dependent critical collective coupling strength. The scattering experiments are performed in the visible spectrum and are known to be sensitive to spatial structuring/clustering of the electronic structure by the molecular ensemble (synchronization and resonance behavior under collective VSC). In addition, the here-proposed entorpically freed electrons  may also contribute significantly to the enhanced scattering intensities. Free electrons undergo Thomson scattering, with a large scattering cross section and thus high intensities per scatterer. It would be interesting to verify and disentangle experimentally these two potential sources (ordering / clustering of molecular ensemble vs. free electrons) of scattering enhancements  for the experimentally measured two order of magnitude increase of scattering intensities, in order to better understand the underlying mechanisms. Currently, we would expect that both effects play a role. An ensemble ordering is likely required to form a collectively degenerate state of the electronic structure, which is required to trigger  transversal electron correlation phase transitions\cite{sidler_collectively-modified_2026,sidler_cavity-mediated_2026}   that may (metallic case) or may not (insulating case) entropically ionize electrons.

To summarize, based on our previously introduced spin glass solution for the collective transversal electron correlation under collective strong coupling, we could derive non-perturbatively the corresponding temperature-dependent static polarizability in the thermodynamic limit. The derived observable shows characteristic features upon entering the previously proposed para- or spin-glass correlated phases.
For the fractionally occupied degenerate case, a locally divergent polarizability confirms the cavity-induced metallic behavior suggested in Ref. \cite{sidler_cavity-mediated_2026}. The metallic behavior emerges from a collective entropy increase that favors the fractional occupation of the collectively degenerate state, whose electrons act similarly to free / ionized electrons.  We believe,  the here-derived polarizability phase-diagrams (Fig. \ref{fig:polarizabilities}) could help understanding different experiments that report strong conductivity / dielectric constant changes\cite{fukushima_inherent_2022,kumar_extraordinary_2024} and temperature-dependent phase transitions under collective strong coupling.\cite{patrahau_direct_2024, sandeep_cluster_2026} So far, we lack any theoretical explanation of these observations. In the future, we plan to extend our theory to connect in the thermodynamic limit with above experimental observables (dielectric constant, conductivity, or scattering intensities) and investigate the impact of collective electron correlation phases on the nuclear dynamics (e.g., self-assembly and molecular aggregation)\cite{sandeep_manipulating_2022,joseph_supramolecular_2021,joseph_consequences_2024}.


\begin{acknowledgments}
We thank Michael Ruggenthaler for fruitful discussions. This work was made possible through the support of the European Research Council (ERC-2024-SYG-101167294, UnMySt). 
\end{acknowledgments}

\begin{appendix}
\section{Appendix}
 
   Using the Slater-Condon rules, one finds
     \begin{widetext}
\begin{eqnarray}
   \langle\Phi^\mathrm{CIS}| \hat{H}^{'} |\Phi^\mathrm{CIS}\rangle &=& -eE_{loc} \Bigg\{s_0^2\sum_{m\in \rm occ} \bra{\phi_m}\hat{z}\ket{\phi_m}+\sum_{m\in \rm occ}\sum_{p\in \rm unocc}(s_m^p)^2\bigg[ \sum_{n\in \rm  occ} \bra{\phi_n}\hat{z}\ket{\phi_n} - \bra{\phi_m}\hat{z}\ket{\phi_m}+\bra{\phi_n}\hat{z}\ket{\phi_n}\bigg]\\
   &&
   \sum_{m\in \rm occ}\sum_{p\in \rm unocc} s_0 s_m^p \bra{\phi_m}\hat{z}\ket{\phi_p} + \sum_{m\neq m^\prime\in \rm occ}\sum_{p\in \rm unocc} s_{m^\prime}^p s_m^p \bra{\phi_{m^\prime}}\hat{z}\ket{\phi_m} + \sum_{m \in \rm occ}\sum_{p \neq p^\prime\in \rm unocc}  s_{m}^{p^\prime} s_m^p\bra{\phi_{p^\prime}}\hat{z}\ket{\phi_p}\Bigg\}\nonumber
   \\
   &\overset{\mu=0}{=}&-eE_{loc} \Bigg\{\sum_{m\in \rm occ}\sum_{p\in \rm unocc} s_0 s_m^p \bra{\phi_m}\hat{z}\ket{\phi_p} + \sum_{m\neq m^\prime\in \rm occ}\sum_{p\in \rm unocc} s_{m^\prime}^p s_m^p \bra{\phi_{m^\prime}}\hat{z}\ket{\phi_m}  \\
   &&+\sum_{m \in \rm occ}\sum_{p \neq p^\prime\in \rm unocc}  s_{m}^{p^\prime} s_m^p\bra{\phi_{p^\prime}}\hat{z}\ket{\phi_p}\Bigg\}\nonumber
   \\
   &=&-E_{loc}\sum_{i\neq j}^{N_{\rm deg}} L_{ij}\sigma_i \sigma_j
   \end{eqnarray}
   \end{widetext}
 where we have dropped the diagonal elements $L_{ii}$. The reason is that one can  write generically $L_{ii}=\mu+\delta_i$ and identify $\mu=e\sum_{m\in \rm occ} \bra{\phi_m}\hat{z}\ket{\phi_m}$ and $\delta_i=-e\bra{\phi_m}\hat{z}\ket{\phi_m}+e\bra{\phi_n}\hat{z}\ket{\phi_n} $ for our case. 
 Consequently, the total Stark energy of the diagonal elements becomes $-E_{loc}\sum_{i}^{N_{\rm deg}} L_{ii}\sigma_i^2=-E_{loc}\mu N+-E_{loc}\sum_{i}^{N_{\rm deg}} \delta_{i}\sigma_i^2$. 
 To calculate the transversal polarizability $\alpha_\perp$, i.e., for $E_{loc}\rightarrow0$, the linear $\mu$-term will not contribute and we discard it. Furthermore, for our CIS case, we notice that $Var(L_{ii})=Var(\delta_i)=Var(L_{ij})$ and hence, we recover the usual SSK setup, where one can show $\sum_{i}^{N_{\rm deg}} L_{ii}\sigma_i^2\rightarrow 0$ for  $N_{\rm deg}\rightarrow\infty$.

\end{appendix}

\bibliography{references}

@article{castagnola_realistic_2025,
	title = {Realistic \textit{{Ab} {Initio}} {Predictions} of {Excimer} {Behavior} under {Collective} {Light}-{Matter} {Strong} {Coupling}},
	volume = {15},
	issn = {2160-3308},
	url = {https://link.aps.org/doi/10.1103/PhysRevX.15.021040},
	doi = {10.1103/PhysRevX.15.021040},
	language = {en},
	number = {2},
	urldate = {2026-08-17},
	journal = {Physical Review X},
	author = {Castagnola, Matteo and Lexander, Marcus T. and Koch, Henrik},
	month = may,
	year = {2025},
	pages = {021040},
}

@article{martinez-martinez_can_2018,
	title = {Can {Ultrastrong} {Coupling} {Change} {Ground}-{State} {Chemical} {Reactions}?},
	volume = {5},
	issn = {2330-4022, 2330-4022},
	url = {https://pubs.acs.org/doi/10.1021/acsphotonics.7b00610},
	doi = {10.1021/acsphotonics.7b00610},
	language = {en},
	number = {1},
	urldate = {2026-08-17},
	journal = {ACS Photonics},
	author = {Martínez-Martínez, Luis A. and Ribeiro, Raphael F. and Campos-González-Angulo, Jorge and Yuen-Zhou, Joel},
	month = jan,
	year = {2018},
	pages = {167--176},
}

@article{schwennicke_when_2025,
	title = {When do molecular polaritons behave like optical filters?},
	volume = {54},
	issn = {0306-0012, 1460-4744},
	url = {https://pubs.rsc.org/cs/article/54/13/6482-6504/845106},
	doi = {10.1039/D4CS01024H},
	language = {en},
	number = {13},
	urldate = {2026-08-17},
	journal = {Chemical Society Reviews},
	author = {Schwennicke, Kai and Koner, Arghadip and Pérez-Sánchez, Juan B. and Xiong, Wei and Giebink, Noel C. and Weichman, Marissa L. and Yuen-Zhou, Joel},
	year = {2025},
	pages = {6482--6504},
}

@article{hertzog_strong_2019,
	title = {Strong light–matter interactions: a new direction within chemistry},
	volume = {48},
	issn = {0306-0012, 1460-4744},
	shorttitle = {Strong light–matter interactions},
	url = {https://pubs.rsc.org/cs/article/48/3/937-961/600631},
	doi = {10.1039/C8CS00193F},
	language = {en},
	number = {3},
	urldate = {2026-08-17},
	journal = {Chemical Society Reviews},
	author = {Hertzog, Manuel and Wang, Mao and Mony, Jürgen and Börjesson, Karl},
	year = {2019},
	pages = {937--961},
}

@article{mondal_macroscopic_2026,
	title = {A macroscopic condensation theory for vibrational strong coupling effects},
	issn = {2041-1723},
	url = {https://www.nature.com/articles/s41467-026-75222-2},
	doi = {10.1038/s41467-026-75222-2},
	language = {en},
	urldate = {2026-08-14},
	journal = {Nature Communications},
	author = {Mondal, M. Elious and Montillo Vega, Sebastian and Huo, Pengfei},
	month = jul,
	year = {2026},
}

@misc{sidler_cavity-mediated_2026,
	title = {Cavity-mediated localization and collective electron correlation phases},
	url = {http://arxiv.org/abs/2605.01551},
	doi = {10.48550/arXiv.2605.01551},
	urldate = {2026-08-10},
	publisher = {arXiv},
	author = {Sidler, Dominik and Greco, Francesco and Ruggenthaler, Michael and Rubio, Angel},
	month = aug,
	year = {2026},
	note = {arXiv:2605.01551 [quant-ph]},
}

@article{sandeep_cluster_2026,
	title = {Cluster {Formation} and {Phase} {Transitions} {Induced} by {Vibrational} {Strong} {Coupling}},
	volume = {138},
	issn = {0044-8249, 1521-3757},
	url = {https://onlinelibrary.wiley.com/doi/10.1002/ange.202516917},
	doi = {10.1002/ange.202516917},
	language = {en},
	number = {1},
	urldate = {2026-05-01},
	journal = {Angewandte Chemie},
	author = {Sandeep, K. and Swaminathan, S. and Jayachandran, A. and Nagarajan, K. and Gautier, J. and Kushida, S. and Chervy, T. and Vergauwe, R.M.A. and Thomas, A. and Ebbesen, T.W.},
	month = jan,
	year = {2026},
	pages = {e16917},
}

@article{sidler_collectively-modified_2026,
	title = {Collectively-{Modified} {Intermolecular} {Electron} {Correlations}: {The} {Connection} of {Polaritonic} {Chemistry} and {Spin} {Glass} {Physics}: {Focus} {Review}},
	volume = {126},
	copyright = {https://creativecommons.org/licenses/by/4.0/},
	issn = {0009-2665, 1520-6890},
	shorttitle = {Collectively-{Modified} {Intermolecular} {Electron} {Correlations}},
	url = {https://pubs.acs.org/doi/10.1021/acs.chemrev.4c00711},
	doi = {10.1021/acs.chemrev.4c00711},
	language = {en},
	number = {1},
	urldate = {2026-05-01},
	journal = {Chemical Reviews},
	author = {Sidler, Dominik and Ruggenthaler, Michael and Rubio, Angel},
	month = jan,
	year = {2026},
	pages = {4--27},
}

@article{thomas_groundstate_2016,
	title = {Ground‐{State} {Chemical} {Reactivity} under {Vibrational} {Coupling} to the {Vacuum} {Electromagnetic} {Field}},
	volume = {128},
	copyright = {http://creativecommons.org/licenses/by-nc-nd/4.0/},
	issn = {0044-8249, 1521-3757},
	url = {https://onlinelibrary.wiley.com/doi/10.1002/ange.201605504},
	doi = {10.1002/ange.201605504},
	language = {en},
	number = {38},
	urldate = {2024-09-05},
	journal = {Angewandte Chemie},
	author = {Thomas, Anoop and George, Jino and Shalabney, Atef and Dryzhakov, Marian and Varma, Sreejith J. and Moran, Joseph and Chervy, Thibault and Zhong, Xiaolan and Devaux, Eloïse and Genet, Cyriaque and Hutchison, James A. and Ebbesen, Thomas W.},
	month = sep,
	year = {2016},
	pages = {11634--11638},
}

@article{horak_analytic_2025,
	title = {Analytic model reveals local molecular polarizability changes induced by collective strong coupling in optical cavities},
	volume = {7},
	issn = {2643-1564},
	url = {https://link.aps.org/doi/10.1103/PhysRevResearch.7.013242},
	doi = {10.1103/PhysRevResearch.7.013242},
	language = {en},
	number = {1},
	urldate = {2025-10-22},
	journal = {Physical Review Research},
	author = {Horak, Jacob and Sidler, Dominik and Schnappinger, Thomas and Huang, Wei-Ming and Ruggenthaler, Michael and Rubio, Angel},
	month = mar,
	year = {2025},
	pages = {013242},
}

@article{kumar_extraordinary_2024,
	title = {Extraordinary {Electrical} {Conductance} through {Amorphous} {Nonconducting} {Polymers} under {Vibrational} {Strong} {Coupling}},
	volume = {146},
	copyright = {https://doi.org/10.15223/policy-029},
	issn = {0002-7863, 1520-5126},
	url = {https://pubs.acs.org/doi/10.1021/jacs.4c03016},
	doi = {10.1021/jacs.4c03016},
	language = {en},
	number = {28},
	urldate = {2025-05-25},
	journal = {Journal of the American Chemical Society},
	author = {Kumar, Sunil and Biswas, Subha and Rashid, Umar and Mony, Kavya S. and Chandrasekharan, Gokul and Mattiotti, Francesco and Vergauwe, Robrecht M. A. and Hagenmuller, David and Kaliginedi, Veerabhadrarao and Thomas, Anoop},
	month = jul,
	year = {2024},
	pages = {18999--19008},
}

@article{joseph_consequences_2024,
	title = {Consequences of {Vibrational} {Strong} {Coupling} on {Supramolecular} {Polymerization} of {Porphyrins}},
	volume = {146},
	copyright = {https://creativecommons.org/licenses/by/4.0/},
	issn = {0002-7863, 1520-5126},
	url = {https://pubs.acs.org/doi/10.1021/jacs.4c02267},
	doi = {10.1021/jacs.4c02267},
	language = {en},
	number = {17},
	urldate = {2025-05-25},
	journal = {Journal of the American Chemical Society},
	author = {Joseph, Kripa and De Waal, Bas and Jansen, Stef A. H. and Van Der Tol, Joost J. B. and Vantomme, Ghislaine and Meijer, E. W.},
	month = may,
	year = {2024},
	pages = {12130--12137},
}

@article{sandeep_manipulating_2022,
	title = {Manipulating the {Self}-{Assembly} of {Phenyleneethynylenes} under {Vibrational} {Strong} {Coupling}},
	volume = {13},
	copyright = {https://doi.org/10.15223/policy-029},
	issn = {1948-7185, 1948-7185},
	url = {https://pubs.acs.org/doi/10.1021/acs.jpclett.1c03893},
	doi = {10.1021/acs.jpclett.1c03893},
	language = {en},
	number = {5},
	urldate = {2025-05-25},
	journal = {The Journal of Physical Chemistry Letters},
	author = {Sandeep, Kulangara and Joseph, Kripa and Gautier, Jérôme and Nagarajan, Kalaivanan and Sujith, Meleppatt and Thomas, K. George and Ebbesen, Thomas W.},
	month = feb,
	year = {2022},
	pages = {1209--1214},
}

@article{joseph_supramolecular_2021,
	title = {Supramolecular {Assembly} of {Conjugated} {Polymers} under {Vibrational} {Strong} {Coupling}},
	volume = {60},
	issn = {1433-7851, 1521-3773},
	url = {https://onlinelibrary.wiley.com/doi/10.1002/anie.202105840},
	doi = {10.1002/anie.202105840},
	language = {en},
	number = {36},
	urldate = {2025-05-22},
	journal = {Angewandte Chemie International Edition},
	author = {Joseph, Kripa and Kushida, Soh and Smarsly, Emanuel and Ihiawakrim, Dris and Thomas, Anoop and Paravicini‐Bagliani, Gian Lorenzo and Nagarajan, Kalaivanan and Vergauwe, Robrecht and Devaux, Eloise and Ersen, Ovidiu and Bunz, Uwe H. F. and Ebbesen, Thomas W.},
	month = sep,
	year = {2021},
	pages = {19665--19670},
}

@article{baik_spherical_2021,
	title = {Spherical {Spin} {Glass} {Model} with {External} {Field}},
	volume = {183},
	issn = {0022-4715, 1572-9613},
	url = {https://link.springer.com/10.1007/s10955-021-02757-7},
	doi = {10.1007/s10955-021-02757-7},
	language = {en},
	number = {2},
	urldate = {2025-02-04},
	journal = {Journal of Statistical Physics},
	author = {Baik, Jinho and Collins-Woodfin, Elizabeth and Le Doussal, Pierre and Wu, Hao},
	month = may,
	year = {2021},
	pages = {31},
}

@misc{castagnola_changes_2024,
	title = {Changes in excimer properties under collective strong coupling},
	url = {http://arxiv.org/abs/2410.22043},
	urldate = {2024-11-14},
	publisher = {arXiv},
	author = {Castagnola, Matteo and Lexander, Marcus T. and Koch, Henrik},
	month = oct,
	year = {2024},
	note = {arXiv:2410.22043 [physics]},
}

@article{ebbesen_hybrid_2016,
	title = {Hybrid {Light}–{Matter} {States} in a {Molecular} and {Material} {Science} {Perspective}},
	volume = {49},
	issn = {0001-4842, 1520-4898},
	url = {https://pubs.acs.org/doi/10.1021/acs.accounts.6b00295},
	doi = {10.1021/acs.accounts.6b00295},
	language = {en},
	number = {11},
	urldate = {2024-09-09},
	journal = {Accounts of Chemical Research},
	author = {Ebbesen, Thomas W.},
	month = nov,
	year = {2016},
	pages = {2403--2412},
}

@article{sidler_polaritonic_2021,
	title = {Polaritonic {Chemistry}: {Collective} {Strong} {Coupling} {Implies} {Strong} {Local} {Modification} of {Chemical} {Properties}},
	volume = {12},
	issn = {1948-7185, 1948-7185},
	shorttitle = {Polaritonic {Chemistry}},
	url = {https://pubs.acs.org/doi/10.1021/acs.jpclett.0c03436},
	doi = {10.1021/acs.jpclett.0c03436},
	language = {en},
	number = {1},
	urldate = {2023-08-07},
	journal = {The Journal of Physical Chemistry Letters},
	author = {Sidler, Dominik and Schäfer, Christian and Ruggenthaler, Michael and Rubio, Angel},
	month = jan,
	year = {2021},
	pages = {508--516},
}

@article{ruggenthaler_understanding_2023,
	title = {Understanding {Polaritonic} {Chemistry} from {Ab} {Initio} {Quantum} {Electrodynamics}},
	volume = {123},
	issn = {0009-2665, 1520-6890},
	url = {https://pubs.acs.org/doi/10.1021/acs.chemrev.2c00788},
	doi = {10.1021/acs.chemrev.2c00788},
	language = {en},
	number = {19},
	urldate = {2023-11-12},
	journal = {Chemical Reviews},
	author = {Ruggenthaler, Michael and Sidler, Dominik and Rubio, Angel},
	month = oct,
	year = {2023},
	pages = {11191--11229},
}

@article{sidler_perspective_2022,
	title = {A perspective on ab initio modeling of polaritonic chemistry: {The} role of non-equilibrium effects and quantum collectivity},
	volume = {156},
	issn = {0021-9606},
	shorttitle = {A perspective on ab initio modeling of polaritonic chemistry},
	url = {https://doi.org/10.1063/5.0094956},
	doi = {10.1063/5.0094956},
	number = {23},
	urldate = {2023-09-22},
	journal = {The Journal of Chemical Physics},
	author = {Sidler, Dominik and Ruggenthaler, Michael and Schäfer, Christian and Ronca, Enrico and Rubio, Angel},
	month = jun,
	year = {2022},
	pages = {230901},
}

@article{jarc_cavity-mediated_2023,
	title = {Cavity-mediated thermal control of metal-to-insulator transition in {1T}-{TaS2}},
	volume = {622},
	issn = {0028-0836, 1476-4687},
	url = {https://www.nature.com/articles/s41586-023-06596-2},
	doi = {10.1038/s41586-023-06596-2},
	language = {en},
	number = {7983},
	urldate = {2024-09-09},
	journal = {Nature},
	author = {Jarc, Giacomo and Mathengattil, Shahla Yasmin and Montanaro, Angela and Giusti, Francesca and Rigoni, Enrico Maria and Sergo, Rudi and Fassioli, Francesca and Winnerl, Stephan and Dal Zilio, Simone and Mihailovic, Dragan and Prelovšek, Peter and Eckstein, Martin and Fausti, Daniele},
	month = oct,
	year = {2023},
	pages = {487--492},
}

@article{thomas_tilting_2019,
	title = {Tilting a ground-state reactivity landscape by vibrational strong coupling},
	volume = {363},
	issn = {0036-8075, 1095-9203},
	url = {https://www.science.org/doi/10.1126/science.aau7742},
	doi = {10.1126/science.aau7742},
	language = {en},
	number = {6427},
	urldate = {2024-09-09},
	journal = {Science},
	author = {Thomas, A. and Lethuillier-Karl, L. and Nagarajan, K. and Vergauwe, R. M. A. and George, J. and Chervy, T. and Shalabney, A. and Devaux, E. and Genet, C. and Moran, J. and Ebbesen, T. W.},
	month = feb,
	year = {2019},
	pages = {615--619},
}

@article{fukushima_inherent_2022,
	title = {Inherent {Promotion} of {Ionic} {Conductivity} via {Collective} {Vibrational} {Strong} {Coupling} of {Water} with the {Vacuum} {Electromagnetic} {Field}},
	volume = {144},
	copyright = {https://doi.org/10.15223/policy-029},
	issn = {0002-7863, 1520-5126},
	url = {https://pubs.acs.org/doi/10.1021/jacs.2c02991},
	doi = {10.1021/jacs.2c02991},
	language = {en},
	number = {27},
	urldate = {2024-09-05},
	journal = {Journal of the American Chemical Society},
	author = {Fukushima, Tomohiro and Yoshimitsu, Soushi and Murakoshi, Kei},
	month = jul,
	year = {2022},
	pages = {12177--12183},
}

@article{sidler_unraveling_2024,
	title = {Unraveling a {Cavity}-{Induced} {Molecular} {Polarization} {Mechanism} from {Collective} {Vibrational} {Strong} {Coupling}},
	volume = {15},
	copyright = {https://creativecommons.org/licenses/by/4.0/},
	issn = {1948-7185, 1948-7185},
	url = {https://pubs.acs.org/doi/10.1021/acs.jpclett.4c00913},
	doi = {10.1021/acs.jpclett.4c00913},
	language = {en},
	number = {19},
	urldate = {2024-09-05},
	journal = {The Journal of Physical Chemistry Letters},
	author = {Sidler, Dominik and Schnappinger, Thomas and Obzhirov, Anatoly and Ruggenthaler, Michael and Kowalewski, Markus and Rubio, Angel},
	month = may,
	year = {2024},
	pages = {5208--5214},
}

@article{xiang_molecular_2024,
	title = {Molecular {Polaritons} for {Chemistry}, {Photonics} and {Quantum} {Technologies}},
	volume = {124},
	copyright = {https://creativecommons.org/licenses/by/4.0/},
	issn = {0009-2665, 1520-6890},
	url = {https://pubs.acs.org/doi/10.1021/acs.chemrev.3c00662},
	doi = {10.1021/acs.chemrev.3c00662},
	language = {en},
	number = {5},
	urldate = {2024-09-05},
	journal = {Chemical Reviews},
	author = {Xiang, Bo and Xiong, Wei},
	month = mar,
	year = {2024},
	pages = {2512--2552},
}

@article{mandal_theoretical_2023,
	title = {Theoretical {Advances} in {Polariton} {Chemistry} and {Molecular} {Cavity} {Quantum} {Electrodynamics}},
	volume = {123},
	copyright = {https://creativecommons.org/licenses/by/4.0/},
	issn = {0009-2665, 1520-6890},
	url = {https://pubs.acs.org/doi/10.1021/acs.chemrev.2c00855},
	doi = {10.1021/acs.chemrev.2c00855},
	language = {en},
	number = {16},
	urldate = {2024-09-05},
	journal = {Chemical Reviews},
	author = {Mandal, Arkajit and Taylor, Michael A.D. and Weight, Braden M. and Koessler, Eric R. and Li, Xinyang and Huo, Pengfei},
	month = aug,
	year = {2023},
	pages = {9786--9879},
}

@article{simpkins_control_2023,
	title = {Control, {Modulation}, and {Analytical} {Descriptions} of {Vibrational} {Strong} {Coupling}},
	volume = {123},
	copyright = {https://doi.org/10.15223/policy-001},
	issn = {0009-2665, 1520-6890},
	url = {https://pubs.acs.org/doi/10.1021/acs.chemrev.2c00774},
	doi = {10.1021/acs.chemrev.2c00774},
	language = {en},
	number = {8},
	urldate = {2024-09-05},
	journal = {Chemical Reviews},
	author = {Simpkins, Blake S. and Dunkelberger, Adam D. and Vurgaftman, Igor},
	month = apr,
	year = {2023},
	pages = {5020--5048},
}

@article{hirai_molecular_2023,
	title = {Molecular {Chemistry} in {Cavity} {Strong} {Coupling}},
	volume = {123},
	copyright = {https://doi.org/10.15223/policy-029},
	issn = {0009-2665, 1520-6890},
	url = {https://pubs.acs.org/doi/10.1021/acs.chemrev.2c00748},
	doi = {10.1021/acs.chemrev.2c00748},
	language = {en},
	number = {13},
	urldate = {2024-09-05},
	journal = {Chemical Reviews},
	author = {Hirai, Kenji and Hutchison, James A. and Uji-i, Hiroshi},
	month = jul,
	year = {2023},
	pages = {8099--8126},
}

@article{bhuyan_rise_2023,
	title = {The {Rise} and {Current} {Status} of {Polaritonic} {Photochemistry} and {Photophysics}},
	volume = {123},
	copyright = {https://creativecommons.org/licenses/by/4.0/},
	issn = {0009-2665, 1520-6890},
	url = {https://pubs.acs.org/doi/10.1021/acs.chemrev.2c00895},
	doi = {10.1021/acs.chemrev.2c00895},
	language = {en},
	number = {18},
	urldate = {2024-09-05},
	journal = {Chemical Reviews},
	author = {Bhuyan, Rahul and Mony, Jürgen and Kotov, Oleg and Castellanos, Gabriel W. and Gómez Rivas, Jaime and Shegai, Timur O. and Börjesson, Karl},
	month = sep,
	year = {2023},
	pages = {10877--10919},
}

@article{fregoni_theoretical_2022,
	title = {Theoretical {Challenges} in {Polaritonic} {Chemistry}},
	volume = {9},
	copyright = {https://creativecommons.org/licenses/by/4.0/},
	issn = {2330-4022, 2330-4022},
	url = {https://pubs.acs.org/doi/10.1021/acsphotonics.1c01749},
	doi = {10.1021/acsphotonics.1c01749},
	language = {en},
	number = {4},
	urldate = {2024-09-05},
	journal = {ACS Photonics},
	author = {Fregoni, Jacopo and Garcia-Vidal, Francisco J. and Feist, Johannes},
	month = apr,
	year = {2022},
	pages = {1096--1107},
}

@article{ruggenthaler_quantum-electrodynamical_2018,
	title = {From a quantum-electrodynamical light–matter description to novel spectroscopies},
	volume = {2},
	issn = {2397-3358},
	url = {https://www.nature.com/articles/s41570-018-0118},
	doi = {10.1038/s41570-018-0118},
	language = {en},
	number = {3},
	urldate = {2024-09-05},
	journal = {Nature Reviews Chemistry},
	author = {Ruggenthaler, Michael and Tancogne-Dejean, Nicolas and Flick, Johannes and Appel, Heiko and Rubio, Angel},
	month = mar,
	year = {2018},
	pages = {0118},
}

@article{ebbesen_introduction_2023,
	title = {Introduction: {Polaritonic} {Chemistry}},
	volume = {123},
	copyright = {https://doi.org/10.15223/policy-001},
	issn = {0009-2665, 1520-6890},
	shorttitle = {Introduction},
	url = {https://pubs.acs.org/doi/10.1021/acs.chemrev.3c00637},
	doi = {10.1021/acs.chemrev.3c00637},
	language = {en},
	number = {21},
	urldate = {2024-09-05},
	journal = {Chemical Reviews},
	author = {Ebbesen, Thomas W. and Rubio, Angel and Scholes, Gregory D.},
	month = nov,
	year = {2023},
	pages = {12037--12038},
}

@article{sherrington_solvable_1975,
	title = {Solvable {Model} of a {Spin}-{Glass}},
	volume = {35},
	copyright = {http://link.aps.org/licenses/aps-default-license},
	issn = {0031-9007},
	url = {https://link.aps.org/doi/10.1103/PhysRevLett.35.1792},
	doi = {10.1103/PhysRevLett.35.1792},
	language = {en},
	number = {26},
	urldate = {2024-05-02},
	journal = {Physical Review Letters},
	author = {Sherrington, David and Kirkpatrick, Scott},
	month = dec,
	year = {1975},
	pages = {1792--1796},
}

@article{patrahau_direct_2024,
	title = {Direct {Observation} of {Polaritonic} {Chemistry} by {Nuclear} {Magnetic} {Resonance} {Spectroscopy}},
	issn = {1433-7851, 1521-3773},
	url = {https://onlinelibrary.wiley.com/doi/10.1002/anie.202401368},
	doi = {10.1002/anie.202401368},
	language = {en},
	urldate = {2024-05-02},
	journal = {Angewandte Chemie International Edition},
	author = {Patrahau, B. and Piejko, M. and Mayer, R. J. and Antheaume, C. and Sangchai, T. and Ragazzon, G. and Jayachandran, A. and Devaux, E. and Genet, C. and Moran, J. and Ebbesen, T. W.},
	month = may,
	year = {2024},
	pages = {e202401368},
}

 \end{document}